# WLAN Specific IoT Enable Power Efficient RAM Design on 40nm FPGA

Tanesh Kumar, Faizan Khan, Safeeullah Soomro, Areez Khalil Memon
Faculty of Engineering, Sciences and Technology
Indus University
Karachi, Pakistan
tanesh.kumar@indus.edu.pk,    faizan76-2011@indus.edu.pk,    ssoomro@indus.edu.pk, areez.memon@indus.edu.pk

*Abstract*— **Increasing the speed of computer is one of the important aspects of the Random Access Memory (RAM) and for better and fast processing it should be efficient. In this work, the main focus is to design energy efficient RAM and it also can be accessed through internet. A 128-bit IPv6 address is added to the RAM in order to control it via internet. Four different types of Low Voltage CMOS (LCVMOS) IO standards are used to make it low power under five different WLAN frequencies is taken. At WLAN frequency 2.4GHz, there is maximum power reduction of 85% is achieved when LVCMOS12 is taken in place of LVCMOS25. This design is implemented using Virtex-6 FPGA, Device xc6vlx75t and Package FF484.**
*Keywords—WLAN, Internet of Things (IoT), RAM, FPGA, LVCMOS, IO Power.*

## I. INTRODUCTION

The modern era of technology has turned on many things, we are going back to the time when computers have a totally different concept about memory, how it is evolved and about its bright future, began with a vacuum tube, got a breakthrough with solid state memory devices in integrated circuits IC, using transistors technology invented at Bell Labs, developing as a core component of advance memory systems, started out with latches, flip flops, capacitors forming registers. Random Access Memory (RAM) ordinarily adds up in a memory chip form, some are microscopic, small indeed to be retain in one's pocket. RAM has fulfilled a significant character in computers, without RAM we won't be capable of doing what computers are now days able to do, advancing as a key component in modern hardware as they developed frequently. RAM truly adding up to the technology in 1966 when Robert Dennard went on with the primary thought for Dynamic Random Access Memory DRAM in IBM's Research Center, arisen as a notable advancement in computers, aspiring enough for Intel to bring out first RAM 1103 in late 1970's, sizing to 1Kbit 125 bytes of memory, setting up a benchmark as the effective semiconductor and occupied by Hewlett Packard (HP) for their computers.

Before this technology, in 1949 An Wang and Way-Dong Woo at Harvard University laboratory invented Core memory, getting popular and cheap indeed to substitute Drum memory and Vacuum tubes technology in 1960's. Modern computer a system runs at least 256 Byte of RAM, while in gaming it usually goes in Gigabyte (i.e. 1 Gb to 6 Gb). Despite other industries driving the business of memory as well but gaming industry play a vital role in boosting up this technology as they still bought huge advancements in graphics, 3d-animations to provide gamer a healthy environment, demanding speedy systems with a lot of computational power for efficient performance. Current Generations are using DDR2 and DDR3 (i.e. Double Data Rate Ram 2) can run upto 800MHz and 1600MHz maximum respectively, bounded upto 4000Mbps in DDR2 and 8000Mbps in DDR3.

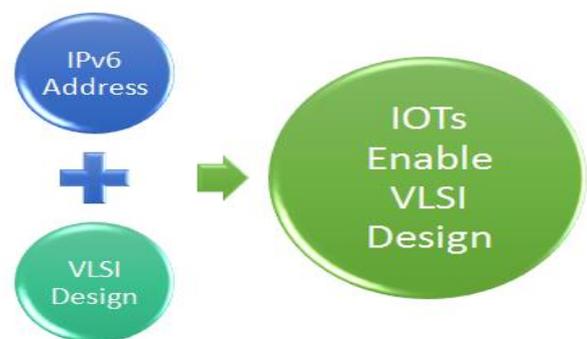

Figure 1: IoTs Enable VLSI Design

In the recent era IPv6 address is added to the targeted VLSI design in order to make internet of things (IOTs) enable VLSI Design as show in figure 1.

Several types of RAMS are utilized today. Dynamic RAM is mostly employed, storing bits of data applying transistors and capacitors coupled as a single cell of memory. Static RAM occupies 4 or more extent of transistors for data storage. Term "Static" refers to constant state and doesn't need to be refreshed while, DRAM must be refreshed or pinged on regular basis to conserve the electric charge on capacitors. Both are volatile, losing the information when power is released. Marketplace, operated via android, iOS and windows phone devices and its information is kept on cloud.

IOT is examined in various FPGA devices to design diverse devices for various applications. As the number of IOT

devices are increasing rapidly, results the need to build a smart world of devices with less human influences. General idea of IOT is to assign an IP address to a relative device to be pinged out from remote distance, widely used for real time data collection. In this work energy efficient RAM is implemented on FPGA and making it IoT enable. IoT Enable RAM is operated under following frequency standard of WLAN (Wireless Local Area Network)

Table 1: List of WLAN Channels [9-12]

| WLAN Channels | Frequency (GHz) | Range and Specification |
|---|---|---|
| 802.11ah | 900 MHz | Unlicensed Bands |
| 802.11b/g/n | 2.4 GHz | 2412-2484MHz |
| 802.11y | 3.6 GHz | 3657.5-3692.5MHz |
| 802.11a/h/j/n/ac | 5 GHz | 4915-5825MHz |
| 802.11p | 5.9 GHz | 5850-5925MHz |

This gives us feasible option to test the reliability of our target device that is RAM. By this we can connect the target device through WLAN frequency channels and perform communication through these channels.

## II. LITERATURE REVIEW

The work in [1] highlight the worthiness of RAM based FPGA testing and proposed a conventional approach relevant to test the structure of RAM based FPGA reporting its configurability deftness while manufacturing. Several approaches with distinct objectives are confronted and then compared with classical buses, outcomes reveals that only three test gets up to the mark, ensuring 100% fault non-redundancy, eventually proposed invariant number of test configurations and short sequences. In [2] research focus on the cryptographic primitives for the security of wireless systems, presents a hardware architecture and RAM based FPGA implementation on FPGA of 64-Bit NESSIE proposal, MISTY1 block cipher, considering eminent throughput essentials, enduring encrypting and decrypting in systems, unfurling of the MISTY1 rounds within a 75-stage pipeline achieving the utmost 12.6 Gbps throughput bearing frequency at 168 MHz. A non-volatile FPGA circuit grounded on Spin-RAM and MRAM technology proposed in [3], having speedy processing and lower power dissipation, Magnetic Tunnel Junction (MTJs) are occupied as storage components. Multi-configuration or dynamic FPGA capable of being implementing with a quiet ease and small-scale surface overhead, resulting to become a technological option in multiprocessing circuits. Spin-RAM technology based Flip-Flop be utilized to substitute the registers in System-on-chip (SOC) to transform these chips secure and non-volatile.

The work on [4] gives detail discussion on the 8bit encryption/decryption hardware execution on block RAM occupying about 130 slice and 4 BRAMs attaining throughput of 27Mbps, equating with ASIP design resulting increase of 8% in slice number while achieving 12 times more throughput. Mix Column could be proceeding to BRAM through which reducing slice numbers might be conceivable. Also discovered that 8-bit AES can additionally be implemented on 32-bit AES in order to reduce slice numbers and to advances frequency. A non-volatile CMOSSTTRAM hybrid FPGA architecture is introduced in [5], recognizing its designing challenges, proposed optimization proficiency at multilevel abstraction. The architecture advantage on high integration density of emerging STTRAM devices and reduces the absolute logic by reducing the individual efforts of CMOS latches, experiencing un-optimized state demanding high power in CMOS counterpart comparatively. An effective resistor-divider design paired with application mapping method stationed on preferential storage facilitate in shortening of the overall power ingestion. Shannon decomposition power gating was implemented to significantly abridge the power dissipation. Results show that non-volatile FPGA will have a promising future in computing. The work in [6] tells about the designing of complex systems and promoting FPGA's as the future technology arise in a very short span of time providing modern designing methods, designing in a new manner, leading to maximum utilization of hardware, have already been the best related technology, having multiple benefits due to its flexible platform. On can find appropriate solution of errors while in design phase, also facilitates in software modeling and emulation. By using FPGA we have derived a new methodology that bus memory system has a bit reflection to superiority of modern systems.

The discussion in [7] shows the implementation of FPGA unprecedented extending chances to implement single chip (DLM) Distributed Logic-Memory architectures effectively with a recognizable hardware path that can calculate precedence among memory partitions, can substantially amend the functioning and energy efficiency of intensive memory applications peculiarly those components that presents irregular access data convention in algorithmic phase. While implementing DLM baffles unique disputes to an FPGA designer in several cases of partitioning and organizing various memory sources, directing effective data transmitting between on and off-chip memory systems. Proposed a hypothetical memory partitioning strategy using annealing algorithm, prevailing partitioning solutions offering parallelized access. The security and privacy issues of wireless sensor systems is highlighted in [8], caused by excess deployment, In order to attain the secrecy of system peculiar cryptographic techniques are used. It can also be practiced in software applying on lower end embedded hardware or in hardware via FPGAs. Author proposed a two distinct FPGA/CPLD implementation of PRESENT (i.e. a lightweight encryption algorithm applicable for Wireless Sensor Networks), obtaining reasonable throughput, notable strategy is to utilize maximum potential of RAM blocks promoting internal states storage, therefore slice counts are reduced. In [13] four different types of low power CMOS (LVCMOS) is implemented for energy efficient frequency meter design on Virtex-6 FPGA Energy efficient finite impulse response (FIR) filter is implemented using thermal aware approach in [14]. In [15] different classes of LVDCI and HSLVDCI Io standards are used to have low power Fibonacci generator design on 40 nm ultra-scale FPGA. Heat Sink, Airflow and Ambient temperature are the important parameters in this mechanism. In [16] low voltage CMOS (LVCMOS) Io standard is used in order to make thermal energy efficient Fibonacci generator design on FPGA and where as in [17] the target design is frame buffer and HSTL IO standard has been implemented to get power efficient design. Performance, energy efficiency and portability of arithmetic logic unit (ALU) is focused in [18]. LVCMOS IO standard is used to achieve the energy efficiency while for the portability and high performance MOBILE DDR IO standard is used.

### III. LOW VOLTAGE CMOS FOR IOT ENABLE RAM

Low Voltage Complementary Metal Oxide Semiconductor (LVCMOS) IO standard is technique used to reduce the power of the electronic devices. In this work four different kinds of LVCMOS has been taken that are LVCMOS12, LVCMOS15, LVCMOS18 and LVCMOS25. The target device is RAM which in IOT enable. The RAM is operated under different WLAN channel frequencies that are 0.9GHz, 2.4GHz, 3.6GHz, 5GHz and 5.9GHz.

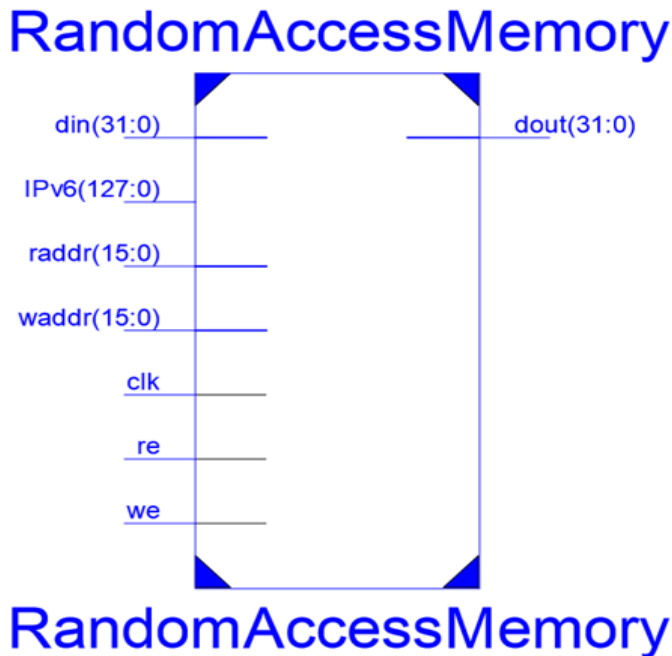

Figure 2: Top Level Schematic of IoT Enable RAM

Figure 2 shows top level schematic of IoT enable RAM which contains 128-bit IPv6 address by which we can control the target device through internet. Other inputs include din, read address (raddr), write address (waddr), clock (clk), read (re) and write (we) and where as output is dout which is of 32-bits.

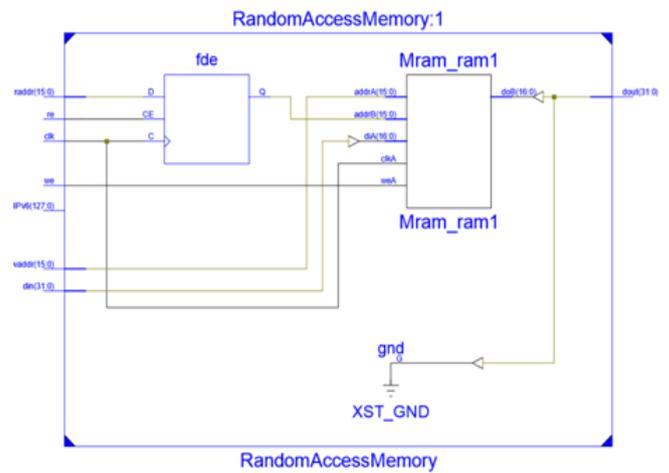

Figure 3: RTL Schematic of IoT Enable RAM

RTL schematic for IoT enable RAM is shown in figure 3 which tells that it contains 16 flip flop/latches, 1 clock buffer. There are 83 input output buffers and out of them 51 are Input buffer and 32 are output buffers.

*A. Power Consumption Using LVCMOS at 0.9GHz*

Table 2: Power Consumption Using LVCMOS at 0.9GHz

|  | LVCMOS 12 | LVCMOS 15 | LVCMOS 18 | LVCMOS 25 |
| --- | --- | --- | --- | --- |
| Clock Power | 0.061W | 0.061W | 0.061W | 0.061W |
| Signal Power | 0.033W | 0.033W | 0.033W | 0.033W |
| BRAM Power | 1.148W | 1.148W | 1.148W | 1.148W |
| IO Power | 0.060W | 0.086W | 0.109W | 0.171W |
| Leakage Power | 1.321W | 1.322W | 1.323W | 1.325W |
| Total Power | 2.624W | 2.651W | 2.675W | 2.739W |

When the RAM is operated at the frequency of 0.9GHz, it is observed that Clock Power, Signal Power and BRAM Power remains constant at each of the type of LVCMOS where IO Power and Leakage Power are not uniform. Leakage power is reduced up to 0.30% reduction while on using of LVCMOS12 in place of LVCMOS25 as shown in table 2 and figure 4.

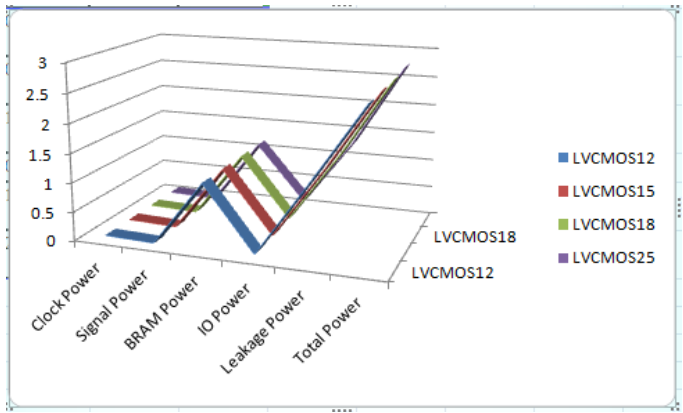

*Figure 4:* Power Consumption Using LVCMOS at 0.9GHz

B. *Power Consumption Using LVCMOS at 2.4GHz*

Table 3: Power Consumption Using LVCMOS at 2.4GHz

|  | LVCMOS 12 | LVCMOS 15 | LVCMOS 18 | LVCMOS25 |
|---|---|---|---|---|
| Clock Power | 0.161W | 0.161W | 0.161W | 0.161W |
| Signal Power | 0.091W | 0.091W | 0.091W | 0.091W |
| BRAM Power | 3.062W | 3.062W | 3.062W | 3.062W |
| IO Power | 0.160W | 0.229W | 0.292W | 0.457W |
| Leakage Power | 1.374W | 1.376W | 1.378W | 1.383W |
| Total Power | 4.849W | 4.920W | 4.985W | 5.155W |

From table 3 and figure 5, it can be exmained that Clock Power, Signal Power and BRAM Power have uniform values on all kind of LVCMOS. Leakage power has less power reduction as compared to IO power. It is analyzed that IO power saved power up to 64.98% on taking LVCMOS12 instead of LVCMOS25 with device operating frequency is taken as 2.4GHz.

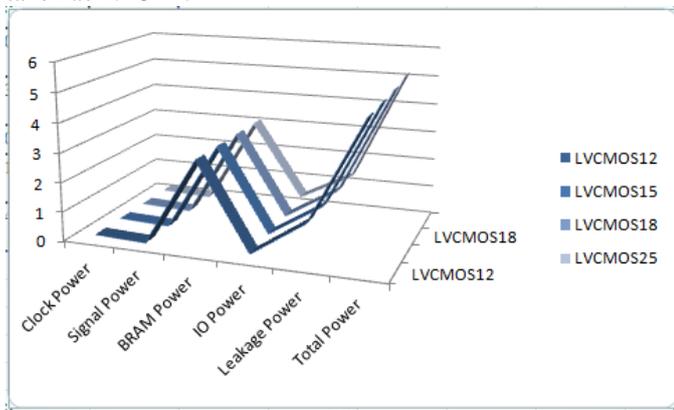

*Figure 5:* Power Consumption Using LVCMOS at 2.4GHz

C. *Power Consumption Using LVCMOS at 3.6GHz*

Table 4: Power Consumption Using LVCMOS at 3.6GHz

|  | LVCMOS 12 | LVCMOS 15 | LVCMOS 18 | LVCMOS25 |
|---|---|---|---|---|
| Clock Power | 0.246W | 0.246W | 0.246W | 0.246W |
| Signal Power | 0.138W | 0.138W | 0.138W | 0.138W |
| BRAM Power | 4.593W | 4.593W | 4.593W | 4.593W |
| IO Power | 0.240W | 0.343W | 0.437W | 0.686W |
| Leakage Power | 1.419W | 1.422W | 1.425W | 1.433W |
| Total Power | 6.637W | 6.744W | 6.841W | 7.096W |

IO power has significant impact on overall power as compared to all other static and dynamic power. At 3.6GHz WLAN frequency and instead of LVCMOS25, LVCOMS12 is taken into consideration; there is 6.46% reduction in total power of the device as shown in table 4 and figure 6.

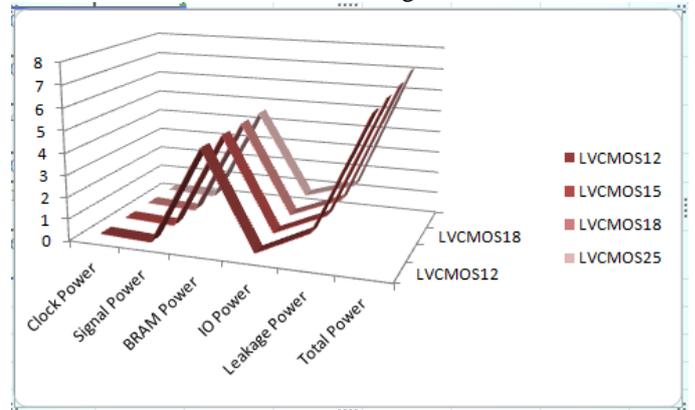

*Figure 6:* Power Consumption Using LVCMOS at 3.6GHz

D. *Power Consumption Using LVCMOS at 5GHz*

Table 5: Power Consumption Using LVCMOS at 5GHz

|  | LVCMOS 12 | LVCMOS 15 | LVCMOS 18 | LVCMOS25 |
|---|---|---|---|---|
| Clock Power | 0.341W | 0.341W | 0.341W | 0.341W |
| Signal Power | 0.192W | 0.192W | 0.192W | 0.192W |
| BRAM Power | 6.380W | 6.380W | 6.380W | 6.380W |
| IO Power | 0.333W | 0.477W | 0.608W | 0.952W |
| Leakage Power | 1.476W | 1.480W | 1.485W | 1.496W |
| Total Power | 8.724W | 8.872W | 9.007W | 9.363W |

Leakage power is static power which has small reduction as compared to IO power and meanwhile it has less impact on overall power of the RAM. As shown in table 5 and figure 7, Leakage power is reduce up to 1.33% in the case when target device is operating at 5.0GHz frequency and LVCMOS12 is taken in place of LVCOMS12.

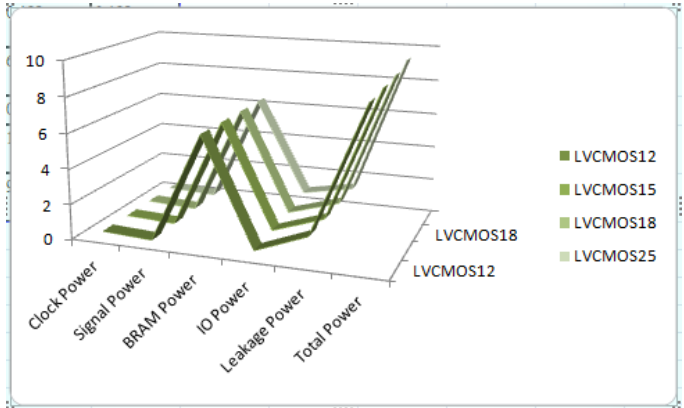

*Figure 7:* Power Consumption Using LVCMOS at 5GHz

### E. Power Consumption Using LVCMOS at 5.9GHz

Table 6: Power Consumption Using LVCMOS at 5.9GHz

|  | LVCMOS12 | LVCMOS15 | LVCMOS18 | LVCMOS25 |
|---|---|---|---|---|
| Clock Power | 0.403W | 0.403W | 0.403W | 0.403W |
| Signal Power | 0.226W | 0.226W | 0.226W | 0.226W |
| BRAM Power | 7.528W | 7.528W | 7.528W | 7.528W |
| IO Power | 0.393W | 0.563W | 0.717W | 1.124W |
| Leakage Power | 1.515W | 1.520W | 1.525W | 1.539W |
| Total Power | 10.067W | 10.242W | 10.402W | 10.822W |

From table 6 and figure 8, it can be analyzed that IO power of the target device RAM is condensed up to 65% when WLAN frequency is taken as 5.9GHz and LVCMOS12 is on higher priority as compared to LVCMOS25.

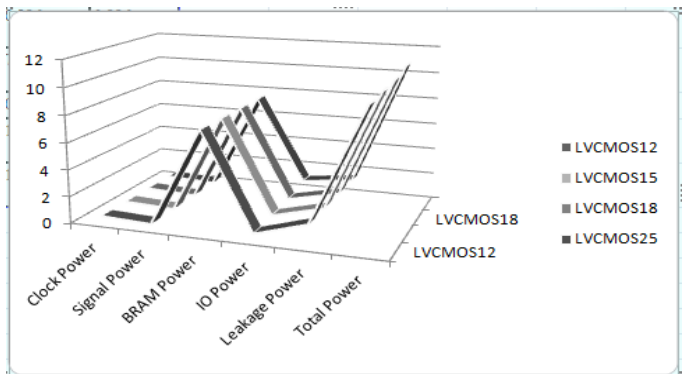

*Figure 8:* Power Consumption Using Capacitance Scaling at 5.9GHz

### F. Comparison of IO Power at Different WLAN frequencies

Table 7: Comparison of Power Reduction at Different WLAN frequencies

|  | 0.9GHz | 2.4GHz | 3.6GHz | 5.0GHz | 5.9GHz |
|---|---|---|---|---|---|
| LVCMOS12 | 0.060 | 0.160 | 0.240 | 0.333 | 0.393 |
| LVCMOS15 | 0.086 | 0.229 | 0.343 | 0.477 | 0.563 |
| LVCMOS18 | 0.109 | 0.292 | 0.437 | 0.608 | 0.717 |
| LVCMOS25 | 0.171 | 1.383 | 0.686 | 0.952 | 1.124 |

As the IO power has the significant impact on total power, due to that it is necessary to examine the IO power of RAM. Table 7 and figure 9 shows the comparison of four available LVCMOS which are driven at different WLAN frequencies. A good amount of power reduction is observed in the target device RAM. It is analyzed that on selecting LVCMOS12 instead of LVCMOS25, there is 64.91%, 88.45%, 65.01%, 65.02% and 65.03% reduction in IO power of the device, when the target device is operated at WLAN frequencies 0.9GHz, 2.4GHz, 3.6GHz, 5.0GHz and 5.9GHz respectively..

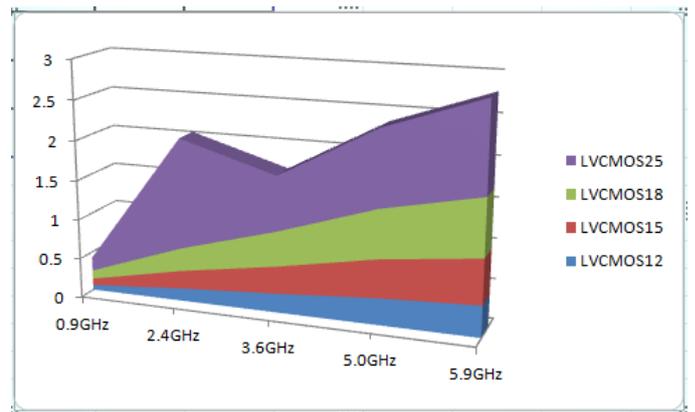

*Figure 8:* Comparison of IO Power at Different WLAN frequencies

## IV. CONCLUSION

This paper deals with design and Implementation of power optimized RAM on 40nm FPGA. This RAM is design as IoTs enable by adding an additional input of 128-bits and through this the RAM can be accessed via internet. Xilinx ISE 14.6 used to simulate this design. Four different types of LVCMOS and five different WLAN frequencies are taken into account. At WLAN frequency 2.4GHz, there is maximum reduction of 85% and at 0.9GHz there is minimum reduction of 64.91% in IO power with using LVCMOS12 instead of LVCMOS25.

## V. FUTURE SCOPE

This design of RAM is implemented using 40nm FPGA, further it can be implemented using 28nm and 16nm ultra scale FPGA. LVCMOS IO standard is used for power efficient design in this work. In future there is a huge scope to reframe se work using other IO standards like Stub Series Terminated Logic (SSTL), High Speed Transceiver Logic (HSTL).

## REFERENCES

[1] M. Renovell, J. Figueras, Y. Zorian, "Test of RAM-Based FPGA: Methodology and Application to the Interconnect", 15th IEEE VLSI Test Symposium, pp. 230-237, 1997, Monterey, CA.

[2] P. Kitsos, M.D. Galanis, and O. Koufopavlou, "A RAM-Based FPGA Implementation of the 64-bit MISTY1 Block Cipher", IEEE International Symposium on Circuits and Systems, ISCAS, pp. 4641 – 4644, 2005.

[3] W. Zhao, E. Belhaire, Q. Mistral, E. Nicolle, T. Devolder, C. Chappert, "Integration of Spin-RAM technology in FPGA circuits", 8th International Conference on Solid-State and Integrated Circuit Technology ICSICT, pp. 799 – 802, 2006, Shanghai.

[4] Chi-Jeng Chang, Chi-Wu Huang, Hung-Yun Tai, Mao-Yuan Lin, Teng-Kuei Hu, "8-bit AES FPGA Implementation using Block RAM", The 33rd Annual Conference of the IEEE Industrial Electronics Society (IECON), pp. 2654-2659 , 2007, Taipei, Taiwan.

[5] Somnath Paul, Saibal Mukhopadhyay, Swarup Bhunia, "Hybrid CMOS-STTRAM Non-Volatile FPGA: Design Challenges and Optimization Approaches", IEEE/ACM International Conference on Computer-Aided Design ICCAD, pp. 589-592, 2008, San Jose, CA.

[6] Yin Dong, Jun Yang, Weiping Zhang, Xiaojun Wang, "The Implementation of Computer Bus-RAM Based on FPGA", 2nd International Conference on Artificial Intelligence, Management Science and Electronic Commerce (AIMSEC), pp. 2554-2557, 2011, Deng Leng.

[7] Yu Bai, Abigail Fuentes, Michael Riera, Mohammed Alawad, Mingjie Lin, "Boosting Memory Performance of Many-Core FPGA Device through Dynamic Precedence Graph", 21st Annual International IEEE Symposium on Field-Programmable Custom Computing Machines, pp.21-24, 2013.

[8] Elif Bilge Kavun, Tolga Yalcin, "RAM-Based Ultra-Lightweight FPGA Implementation of PRESENT", International Conference on Reconfigurable Computing and FPGAs, pp. 280-285, 2011, Cancun.

[9] Halperin, Daniel, et al. "Predictable 802.11 packet delivery from wireless channel measurements." ACM SIGCOMM Computer Communication Review 41.4 (2011): 159-170.

[10] Tsao, Shiao-Li, and Chung-Huei Huang. "A survey of energy efficient MAC protocols for IEEE 802.11 WLAN." Computer Communications 34.1 (2011): 54-67.

[11] Ong, Eng Hwee, et al. "IEEE 802.11 ac: Enhancements for very high throughput WLANs." IEEE 22nd International Symposium on Personal Indoor and Mobile Radio Communications (PIMRC), 2011.

[12] IEEE 802.11™: Wireless LANs, IEEE Standard for Information technology--Telecommunications and information exchange http://standards.ieee.org/about/get/802/802.11.html

[13] Tanesh Kumar, Bishwajeet Pandey, and Teerath Das, "LVCMOS 1/0 Standard And Drive Strength Based Green Design on Ultra Scale FPGA", IEEE International conference on Green Computing, Communication and Conservation of Energy(ICGCE), 12-14 December, 2013.

[14] B. Pandey, T.Das, T.Kumar, J. Kumar, "Thermal Mechanics based Energy Efficient FIR filter for Digital Signal Processing," Applied Mechanics and Materials(AMM) Journal, ISSN:1662-7482(online version).

[15] T.Kumar, B. Pandey, T. Das. "Digitally controlled impedance based green design on ultra scale FPGA." IEEE International Conference on Green Computing, Communication and Conservation of Energy (ICGCE), 2013

[16] Kumar, T., Das, T., Pandey, B., Hussain, D.M.A. (2014). IO Standard Based Thermal/Energy Efficient Green Communication For Wi-Fi Protected Access on FPGA. 6th International Congress on Ultra-Modern Telecommunications and Control systems and Workshops, St. Petersburg, Russia (In Press)

[17] Kumar T., Limbu M.M., Kumar A., Pandey B., and Das T. (2014). Simulation of HSTL IO Standard Based Energy Efficient Frame Buffer For Digital Image Processor. IEEE International Conference on Robotics & Emerging Allied Technologies in Engineering (iCREATE), National University of Sciences and Technology (NUST), Islamabad, Pakistan.

[18] Kumar, T., Pandey, B., Das, T., & Chowdhry, B. S. (2014). Mobile DDR IO Standard Based High Performance Energy Efficient Portable ALU Design on FPGA. Wireless Personal Communications, 76(3), 569-578.